\documentclass[%
 reprint,
amsmath,amssymb,aps,prd,
longbibliography,
]{revtex4-2}

\usepackage[normalem]{ulem}

\usepackage{xcolor}
\usepackage{physics}
\usepackage{mathtools}
\usepackage{graphicx}
\usepackage{dcolumn}
\usepackage{bm}

\usepackage{booktabs}
\usepackage{tikz}
\usetikzlibrary{arrows.meta, decorations.markings, calc}

\newcommand{\W}{\mathcal{W}}

\newcommand{\SU}[1]{{\mathrm{SU(#1)}}}

\newcommand{\figref}[1]{{FIG.~\ref{#1}}}
\begin{document}

\preprint{APS/123-QED}

\title{{Non-Perturbative Trivializing Flows for Lattice Gauge Theories}}

\author{Mathis Gerdes}
\email{m.gerdes@uva.nl}
\affiliation{%
 Institute of Physics, University of Amsterdam
}%
\author{Pim de Haan}%
\affiliation{%
 CuspAI
}%
\author{Roberto Bondesan}
\affiliation{
 Department of Computing, Imperial College London
}%
\author{Miranda C. N. Cheng}
\altaffiliation[On leave from ]{CNRS, France.}%
\affiliation{%
Institute of Physics, University of Amsterdam}%
\affiliation{%
Institute for Mathematics, Academia Sinica, Taiwan} %
\affiliation{%
Korteweg-de Vries Institute for Mathematics, University of Amsterdam
}%

\begin{abstract}
    Continuous normalizing flows are known to be highly expressive and flexible, which allows for easier incorporation of large symmetries and makes them a powerful computational tool for lattice field theories. Building on previous work, we present a general continuous normalizing flow architecture for matrix Lie groups that is equivariant under group transformations. We apply this to lattice gauge theories in two dimensions as a proof of principle and demonstrate competitive performance, showing its potential as a tool for future lattice computations.
\end{abstract}

\maketitle

\tableofcontents

\section{Introduction}

Lattice gauge theory is a cornerstone of theoretical physics, providing non-perturbative insights into quantum field theories such as Quantum Chromodynamics (QCD).
Traditional computational methods like Markov Chain Monte Carlo (MCMC) \cite{morningstar2007MonteCarloa} suffer from critical slowing down near phase transitions and in the continuum limit, leading to significant computational challenges \cite{sokal1997monte}.

Normalizing flows are machine learning models that produce samples from a target distribution by learning a bijective transformation between an easy-to-sample distribution and the target \cite{papamakarios2021normalizing}.
They have emerged as a fundamental approach to computing observables in lattice field theory,
promising to accelerate computations by improving sampling efficiency and reducing autocorrelation times \cite{albergo2019FlowbasedGenerative}. See \cite{albergo2021IntroductionNormalizing,nikiLect} for an overview.

Explicitly preserving symmetries is crucial in designing efficient samplers based on normalizing flows.
This is because symmetries allow one to restrict the class of possible bijective transformations to be learned, thereby reducing the sample complexity of the task.
The importance of symmetries is especially pronounced when simulating lattice gauge theories, where the size of the gauge symmetry group grows rapidly with the lattice size.
Given their high design flexibility, continuous-time normalizing flows \cite{chen2019neuralordinarydifferentialequations} are a promising framework.
This is to be contrasted with discrete flows \cite{albergo2019FlowbasedGenerative, boyda2021SamplingUsing}, where a partition of the lattice, with a partial breaking of the translation symmetry as a result, is required.
For instance, in \cite{dehaan2021ScalingMachine,gerdes2023LearningLattice} continuous-time flows equivariant to spatial symmetries have been developed for scalar field theories with $\phi^4$ coupling, reaching higher performance than their discrete-time counterpart.
Other contexts in which continuous normalizing flows play a role in physics sampling include \cite{Caselle_2024}.
In the following, we shall simply refer to continuous-time flows as continuous flows and discrete-time flows as discrete flows.

For gauge theories, discrete normalizing flows, such as those employing residual networks {\cite{Nagai:2021bhh,abbott2023NormalizingFlows,abbott2024applicationModels}} and coupling layers \cite{boyda2021SamplingUsing}, have been developed to respect gauge symmetries.
In view of the above-mentioned benefits, it is desirable to explore continuous realizations of normalizing flows for lattice gauge theories.  Exactly such a method was theoretically proposed by
Lüscher in the name of ``trivializing maps" for gauge theories in the foundational work
\cite{luscher2010TrivializingMaps}, before the advent of the modern machine learning technology.
In particular, the Lüscher flow is driven by a potential that is assumed to admit a perturbative expansion in flow time.
The specific Ansatz of the Lüscher flow then dictates that this perturbative expansion in flow time is also a perturbative expansion
in the inverse coupling constant $\beta$ (see \eqref{eq:wilson-action}). In particular, it has the form $\tilde S =  \beta\sum_k ({\beta t \over 6})^{k} \tilde S^{(k)}$, where $\tilde S^{(k)}$ is independent of $\beta$.
Following this idea, a machine learning-enhanced implementation of such a trivializing flow was recently reported in \cite{bacchio2023LearningTrivializing}, showing promising results in sampling quality.
While effective, this approach did not make full use of the general expressive power of neural networks, constraining itself instead to a specific form of expansion closely following Lüscher's original idea.

Building on these developments, we present a flexible and fully equivariant framework for continuous normalizing flows for lattice gauge theories. The goal of continuous flows is to learn the vector field driving the field transformation such that the endpoint of the flow gives a good approximation of the target density, in this case given in terms of the action.
In contrast to the original Ansatz of Lüscher's, the time dependence of the vector field in our flow is not constrained to be polynomial.
Moreover, while retaining equivariance, we do not constrain the vector field to be given by the gradient of a scalar potential.
Most importantly, our framework allows for general neural network architectures parameterizing the vector field, leading to more expressive machine-learning models. As a result,
the vector field can in principle be an arbitrary function of the field configuration, limited only by the particular choice of architecture which has to be made in consideration of computational efficiency.

In particular, our framework includes the flow of Lüscher \cite{luscher2010TrivializingMaps} and its machine learning implementation \cite{bacchio2023LearningTrivializing}  as special cases.
The additional flexibility of our framework gives us the freedom to choose a more or less complex neural network, as appropriate for the given application,  balancing the trade-off between expressiveness and computational cost.

A feature of neural ODE allowing for efficient gradient computations is the adjoint sensitivity method, which we implemented in a general way.
This approach allows for automatic differentiation and memory-efficient gradient computation for an arbitrary neural network architecture.
This is combined with Crouch-Grossman Runge-Kutta integration schemes that can simultaneously process scalar and matrix group-valued degrees of freedom, ensuring numerical stability.
Our implementation can be straightforwardly applied to any related integration problem involving real numbers and matrix groups, and any loss functions.

We used this framework to construct a novel continuous normalizing flow model for lattice gauge theories. Our method achieves state-of-the-art effective sample size -- a metric quantifying the efficiency of the sampler --
when benchmarked on sampling two-dimensional $\SU{2}$ and $\SU{3}$ lattice gauge theories.
The integration and Lie group infrastructure developed for this study forms part of the broader flow library bijx \cite{Gerdes2025Bijx}.
Numerical data for the figures can be found at \cite{zenodoData}.

\section{Gauge Equivariant Continuous Flows}

The aim of normalizing flows is to approximate samples from a target distribution, which in the present context is given in terms of the actions $S$ as
\begin{equation}
    p(U) = \frac{e^{-S(U)}}{Z} \,,
\end{equation}
where the partition function $Z$ is a normalization constant.
$U$ is a random variable that for lattice gauge theories takes value in $\SU{N}^{|E|}$ with the number of edges is given by $|E| = L^d d$ for a periodic square lattice with side length $L$ in each of the $d$ dimensions.
To sample from $p$, we initially draw samples ${{U \sim \rho(U)}}$ from an easy-to-sample distribution, which we choose that associated to be the Haar measure $DU$.
One then applies a bijective map $U \mapsto U'=\Phi(U)$ to obtain the output samples.
This change of variables induces a probability distribution $q(U')$ on the proposed samples $U'$.
To train the normalizing flow, crucially without having reliable samples from the target distribution, we minimize the reverse Kullback-Leibler (KL) divergence
\begin{equation}
    \int {DU}\,q(U) \log\left(\frac{q(U)}{p(U)}\right) = \log Z + {\mathbb E}_{U \sim q} L(U) \,,
\end{equation}
where
\begin{align}
\label{eq:L}
L(U) = \log q(U) + S(U)
\end{align}
which we refer to as our loss function.
Note that the partition function is a constant, and therefore drops out when computing the gradient with respect to the network parameters, and that we only need to generate samples from the model $q$ and not the target distribution $p$.
In practice, we will not be able to obtain a global minimum of the loss. We use the flow as the proposal distribution in a Metropolis-Hasting algorithm that guarantees asymptotic exactness of the sampling process\cite{albergo2019FlowbasedGenerative},  in importance sampling \cite{ Nicoli:2020njz}, or in conjunction with Hamiltonian Monte Carlo \cite{Albandea:2023wgd}.

Continuous normalizing flows define the map $\Phi$ as the solution to an ordinary differential equation (ODE), whose vector field is given by a neural network.
In contrast to discrete flows, any sufficiently regular ODE is always invertible, and the change of density can be easily computed if the divergence of the vector field is numerically tractable.
This approach leads to highly expressive architectures that can be made to manifestly respect the symmetries of the target theory, leading to higher parameter efficiency.

In the following, we will present architectures for gauge equivariant normalizing flows that achieve state-of-the-art performance on two-dimensional pure gauge theory for both $\SU{2}$ and $\SU{3}$.
We will first introduce the necessary notations and background in the following section.
For simplicity, we will initially suppress the edge index and consider a single copy of $\SU{N}$ (or a matrix Lie group $G$ more generally), as the statements can be easily generalized to the lattice setup.

\subsection{Ordinary Differential Equations on Lie Groups}

Let $G$ be a matrix Lie group and $\mathfrak g$ the corresponding Lie algebra. In the following, we will focus on the case $G=\SU{N}$, but most of the results straightforwardly apply to any matrix Lie group.

Any tangent vector $V \in T_UG$ at the point $U\in G$ can be decomposed in terms of a Lie algebra element $Z\in \mathfrak g$ as $V = Z U$.
Consequently, a neural ODE on $G$ is defined in terms of a neural network $Z_\theta: G \times \mathbb{R} \rightarrow \mathfrak g$, with parameters $\theta$, as
\begin{equation}\label{eqn:ODE1}
  \dot{U}(t) = Z_\theta(U, t) U(t) \,.
\end{equation}

The Lie algebra $\mathfrak g={\mathfrak{su}}(N)$ is spanned by a set of basis vectors $\qty{T_a}_{a=1}^{N^2-1}$, as $\SU{N}$ has dimensionality $N^2-1$.
For instance, for $G=\SU{2}$ we use $T_a = i \sigma_a$, where $\sigma_a$ are the Pauli matrices, while for $G=\SU{3}$ we use $T_a = i \lambda_a$, where $\lambda_a$ are the Gell-Mann matrices.
Requiring these vectors to form an orthonormal basis, the metric is fixed to have the normalization
\begin{equation} \label{eq:scalar-product}
    \langle A, B \rangle = -tr(A B) / 2 \,.
\end{equation}
Correspondingly, a basis for
tangent space $T_UG$ at the point $U$ is given by $\qty{T_a U}_a$.

We can define the directional derivative of a function $f: G \to \mathbb{R}$ as
\begin{equation}
  \partial_a f (U) = \dv{t}\Big|_{t=0} \, f\qty(e^{t T_a} U) \,.
\end{equation}
The gradient of $f$, which yields a tangent vector at $U$, is then given as
\begin{equation}
  \nabla f (U) = \sum_a \partial_a f(U) \, T_a U \,.
\end{equation}

Consider now the situation when the initial point $U(0)\in G$ is sampled from a certain distribution $q_0$ on $G$, and evolves following the ODE of equation \eqref{eqn:ODE1} up to time $T$. The resulting distribution $q_T$ of $U(T)$, incorporating the change of density induced by the flow, also solves an ODE.

Denoting the log-likelihood as $\mathcal{L}(t, U) = \log q_t(U)$ and writing the right-hand side of the ODE \eqref{eqn:ODE1} as $Z_\theta(U, t) U(t)  = \sum_a Z_\theta^a(t, U) T_a U(t)$, we have \cite{chen2018NeuralOrdinary,falorsi2021ContinuousNormalizing}
\begin{equation}
    \mathcal{L}(T, U) = \mathcal{L}(0, U) - \int_0^T \nabla \cdot  \dot{U}(t, U) \dd{t} \,,
\end{equation}
where the divergence can be computed using the Lie algebra basis as
\begin{equation}
    \nabla \cdot \dot{U}(t, U) =
    \sum_a
    \partial_a Z_\theta^a(t, U) \,.
\end{equation}
In the form of an ODE, this is
\begin{equation}
\label{eqn:loglikelihoodODE}
    \dot{\mathcal{L}} = -  \nabla \cdot \dot{U}(t, U) \,.
\end{equation}
In the application in lattice gauge theories, $U$ will be replaced by a collection of group elements $\{U_e\}_e$, one for each edge of the lattice.

\begin{figure*}[tb]
    \centering
    \begin{tikzpicture}[scale=0.8]

\def\xshiftA{2cm} 
\def\xshiftB{3cm} 
\begin{scope}[shift={(0*\xshiftA,0cm)}]
\coordinate (A0) at (0,0);
\coordinate (B0) at (1,0);
\coordinate (C0) at (1,1);
\coordinate (D0) at (0,1);
\draw [red, thick, postaction={decorate}, decoration={
    markings, mark=at position 0.5 with {\arrow{Stealth}}}] (A0) -- (B0);
\draw [red, thick, postaction={decorate}, decoration={
    markings, mark=at position 0.5 with {\arrow{Stealth}}}] (B0) -- (C0);
\draw [red, thick, postaction={decorate}, decoration={
    markings, mark=at position 0.5 with {\arrow{Stealth}}}] (C0) -- (D0);
\draw [red, thick, postaction={decorate}, decoration={
    markings, mark=at position 0.5 with {\arrow{Stealth}}}] (D0) -- (A0);
\filldraw [black] (0.5,0.5) circle (2pt);
\node at (0.5,-0.5) {(0)};  
\end{scope}

\begin{scope}[shift={(1*\xshiftA,0cm)}]  
\coordinate (A1) at (0,0);
\coordinate (D1) at (0,2);
\coordinate (C1) at (1,2);
\coordinate (B1) at (1,0);
\draw [red, thick, postaction={decorate}, decoration={
    markings, mark=at position 0.5 with {\arrow{Stealth}}}] (A1) -- (B1);
\draw [red, thick, postaction={decorate}, decoration={
    markings, mark=at position 0.5 with {\arrow{Stealth}}}] (B1) -- (C1);
\draw [red, thick, postaction={decorate}, decoration={
    markings, mark=at position 0.5 with {\arrow{Stealth}}}] (C1) -- (D1);
\draw [red, thick, postaction={decorate}, decoration={
    markings, mark=at position 0.5 with {\arrow{Stealth}}}] (D1) -- (A1);
\filldraw [black] (0.5, 0.5) circle (2pt);  
\node at (0.5,-0.5) {(1)};  
\end{scope}

\begin{scope}[shift={(2*\xshiftA,0cm)}]  
\coordinate (A2) at (0,0);   
\coordinate (B2) at (1,0);   
\coordinate (C2) at (1,0.9); 
\coordinate (D2) at (0,1.1); 
\coordinate (E2) at (0,2);   
\coordinate (F2) at (1,2);   
\coordinate (G2) at (1,1.1); 
\coordinate (H2) at (0,0.9); 
\coordinate (I2) at (0,0);   

\draw [black, thick, postaction={decorate}, decoration={
    markings, mark=at position 0.5 with {\arrow{Stealth}}}] (A2) -- (B2);
\draw [black, thick, postaction={decorate}, decoration={
    markings, mark=at position 0.5 with {\arrow{Stealth}}}] (B2) -- (C2);
\draw [black, thick, postaction={decorate}, decoration={
    markings, mark=at position 0.5 with {\arrow{Stealth}}}] (C2) -- (D2);
\draw [black, thick, postaction={decorate}, decoration={
    markings, mark=at position 0.5 with {\arrow{Stealth}}}] (D2) -- (E2);
\draw [black, thick, postaction={decorate}, decoration={
    markings, mark=at position 0.5 with {\arrow{Stealth}}}] (E2) -- (F2);
\draw [black, thick, postaction={decorate}, decoration={
    markings, mark=at position 0.5 with {\arrow{Stealth}}}] (F2) -- (G2);
\draw [black, thick, postaction={decorate}, decoration={
    markings, mark=at position 0.5 with {\arrow{Stealth}}}] (G2) -- (H2);
\draw [black, thick, postaction={decorate}, decoration={
    markings, mark=at position 0.5 with {\arrow{Stealth}}}] (H2) -- (I2);

\filldraw [black] (0.5, 0.5) circle (2pt);
\node at (0.5,-.2) [anchor=north] {(2)};  
\end{scope}

\begin{scope}[shift={(3*\xshiftA,0cm)}]  
\coordinate (A3) at (0,0);
\coordinate (B3) at (2,0);
\coordinate (C3) at (2,2);
\coordinate (D3) at (0,2);
\draw [thick, postaction={decorate}, decoration={
    markings, mark=at position 0.5 with {\arrow{Stealth}}}] (A3) -- (B3);
\draw [thick, postaction={decorate}, decoration={
    markings, mark=at position 0.5 with {\arrow{Stealth}}}] (B3) -- (C3);
\draw [thick, postaction={decorate}, decoration={
    markings, mark=at position 0.5 with {\arrow{Stealth}}}] (C3) -- (D3);
\draw [thick, postaction={decorate}, decoration={
    markings, mark=at position 0.5 with {\arrow{Stealth}}}] (D3) -- (A3);
\filldraw [black] (0.5,0.5) circle (2pt);  
\node at (1,-0.5) {(3)};  
\end{scope}

\begin{scope}[shift={(3*\xshiftA+1*\xshiftB,0cm)}]
\coordinate (A4) at (0,0);
\coordinate (B4) at (2,0);
\coordinate (C4) at (2,1);
\coordinate (D4) at (1,1);
\coordinate (E4) at (1,2);
\coordinate (F4) at (0,2);
\draw [thick, postaction={decorate}, decoration={
    markings, mark=at position 0.5 with {\arrow{Stealth}}}] (A4) -- (B4);
\draw [thick, postaction={decorate}, decoration={
    markings, mark=at position 0.5 with {\arrow{Stealth}}}] (B4) -- (C4);
\draw [thick, postaction={decorate}, decoration={
    markings, mark=at position 0.5 with {\arrow{Stealth}}}] (C4) -- (D4);
\draw [thick, postaction={decorate}, decoration={
    markings, mark=at position 0.5 with {\arrow{Stealth}}}] (D4) -- (E4);
\draw [thick, postaction={decorate}, decoration={
    markings, mark=at position 0.5 with {\arrow{Stealth}}}] (E4) -- (F4);
\draw [thick, postaction={decorate}, decoration={
    markings, mark=at position 0.5 with {\arrow{Stealth}}}] (F4) -- (A4);
\filldraw [black] (0.5,0.5) circle (2pt);
\node at (1,-0.5) {(4)};  
\end{scope}

\begin{scope}[shift={(3*\xshiftA+2*\xshiftB,0cm)}]
\coordinate (A5) at (0,0);
\coordinate (B5) at (1,0);
\coordinate (C5) at (1.02, 0.98);
\coordinate (D5) at (0,1);
\coordinate (E5) at (0.98,1.02);
\coordinate (F5) at (2,1);
\coordinate (G5) at (2,2);
\coordinate (H5) at (1,2);
\draw [thick, postaction={decorate}, decoration={
    markings, mark=at position 0.5 with {\arrow{Stealth}}}] (A5) -- (B5);
\draw [thick, postaction={decorate}, decoration={
    markings, mark=at position 0.5 with {\arrow{Stealth}}}] (B5) -- (C5);
\draw [thick, postaction={decorate}, decoration={
    markings, mark=at position 0.5 with {\arrow{Stealth}}}] (E5) -- (D5);
\draw [thick, postaction={decorate}, decoration={
    markings, mark=at position 0.5 with {\arrow{Stealth}}}] (D5) -- (A5);
\draw [thick, postaction={decorate}, decoration={
    markings, mark=at position 0.5 with {\arrow{Stealth}}}] (C5) -- (F5);
\draw [thick, postaction={decorate}, decoration={
    markings, mark=at position 0.5 with {\arrow{Stealth}}}] (F5) -- (G5);
\draw [thick, postaction={decorate}, decoration={
    markings, mark=at position 0.5 with {\arrow{Stealth}}}] (G5) -- (H5);
\draw [thick, postaction={decorate}, decoration={
    markings, mark=at position 0.5 with {\arrow{Stealth}}}] (H5) -- (E5);
\filldraw [black] (0.5,0.5) circle (2pt);
\node at (1,-0.5) {(5)};
\end{scope}

\begin{scope}[shift={(3*\xshiftA+3*\xshiftB,0cm)}]
    \coordinate (A6) at (0,0);
    \coordinate (B6) at (1,0);
    \coordinate (C6) at (1,1); 
    \coordinate (D6) at (0,1);
    \coordinate (F6) at (2,1);
    \coordinate (G6) at (2,2);
    \coordinate (H6) at (1,2);
    
    \def\gap{0.1cm}
    
    \coordinate (I6) at ($(C6) + (\gap, 0)$);  
    \coordinate (J6) at ($(C6) + (-\gap, 0)$); 
    
    \draw [thick, postaction={decorate}, decoration={
        markings, mark=at position 0.5 with {\arrow{Stealth}}}] (A6) -- (B6);
    \draw [thick, postaction={decorate}, decoration={
        markings, mark=at position 0.5 with {\arrow{Stealth}}}] (B6) -- (C6);
    \draw [thick, postaction={decorate}, decoration={
        markings, mark=at position 0.5 with {\arrow{Stealth}}}] (J6) -- (D6);
    \draw [thick, postaction={decorate}, decoration={
        markings, mark=at position 0.5 with {\arrow{Stealth}}}] (D6) -- (A6);

    \draw [thick, postaction={decorate}, decoration={
        markings, mark=at position 0.5 with {\arrow{Stealth}}}] (C6) -- (H6);

    \draw [thick, postaction={decorate}, decoration={
        markings, mark=at position 0.5 with {\arrow{Stealth}}}] (H6) -- (G6);
    \draw [thick, postaction={decorate}, decoration={
        markings, mark=at position 0.5 with {\arrow{Stealth}}}] (G6) -- (F6);
    
    \draw [thick, postaction={decorate}, decoration={
        markings, mark=at position 0.5 with {\arrow{Stealth}}}] (F6) -- (I6);
    
    \filldraw [black] (0.5,0.5) circle (2pt);
    \node at (1,-0.5) {(6)};
\end{scope}

\begin{scope}[shift={(3*\xshiftA+4*\xshiftB,0cm)}]
\coordinate (A7) at (0,0);
\coordinate (D7) at (0,3);
\coordinate (C7) at (1,3);
\coordinate (B7) at (1,0);
\draw [thick, postaction={decorate}, decoration={
    markings, mark=at position 0.5 with {\arrow{Stealth}}}] (A7) -- (B7);
\draw [thick, postaction={decorate}, decoration={
    markings, mark=at position 0.5 with {\arrow{Stealth}}}] (B7) -- (C7);
\draw [thick, postaction={decorate}, decoration={
    markings, mark=at position 0.5 with {\arrow{Stealth}}}] (C7) -- (D7);
\draw [thick, postaction={decorate}, decoration={
    markings, mark=at position 0.5 with {\arrow{Stealth}}}] (D7) -- (A7);
\filldraw [black] (0.5,0.5) circle (2pt);
\node at (0.5,-0.5) {(7)};
\end{scope}

\end{tikzpicture}
    \caption{Loops used in the ODE architecture. The first two loops, highlighted in red, are used for the non-linear superposition function. The anchor in the dual lattice is displayed as a dot.}
    \label{fig:loops}
\end{figure*}

\subsection{Adjoint Sensitivity Method}

The adjoint sensitivity method is a way of computing gradients through the solution of an ODE.
Below we will give a general version of the adjoint sensitivity method.
The generality of the mathematical equations \eqref{eqn:gradient_general} and \eqref{eqn:adjoint_general}  here is reflected in our implementation of the method,
which can be used for any loss and on any manifold as long as it can be embedded in $\mathbb{C}^n$.
We also record the explicit form this takes in equations
\eqref{eqn:adjoint_Lie} and \eqref{eqn:gradient_Lie} for the special case of integrating on matrix Lie groups with the KL divergence loss, as they are relevant for the lattice gauge theory application in the context of normalizing flows.
Further details and explanations as well as the derivation of the specialized expressions can be found in the Appendix \ref{app:adjoint},
and alternative treatments can be found in \cite{bacchio2023LearningTrivializing, lou2020NeuralManifold, falorsi2020neuralordinarydifferentialequations}.

For a general flow on a manifold $\mathcal{M}$ given by the ODE $\dot{z} = f_\theta(z)$, $z\in {\cal M}$,
the gradient of the loss $L(z(T))$ given a fixed initial value $z(0)$ can be computed using
\begin{equation} \label{eqn:gradient_general}
    d_\theta L(z(T)) = - \int_T^0 a(t) \circ d_\theta f_\theta(t, z(t)) \dd{t} \,,
\end{equation}
by integrating backwards from $T$.
This uses the auxiliary \textit{adjoint state} $a(t)$, which is formally a cotangent vector of $\mathcal M$. It is initialized as $a(T) = d_zL\big|_{z(T)}$ and solved for simultaneously using
\begin{equation} \label{eqn:adjoint_general}
\begin{aligned}
  \dot{a}(t) = - a(t) \circ d_z f_\theta\big|_{z(t)} \,.
\end{aligned}
\end{equation}
Here, we have used the more abstract notation from differential geometry. See for instance \cite{nakahara2003GeometryTopology} for an introduction for physicists.
In particular, for $\varphi: {\cal M}\to  {\cal M}'$,  $d\varphi$ denotes the corresponding differential map
$d\varphi_x: T_x{\cal M}\to T_{\varphi(x)}{\cal M}'$ on the tangent spaces, ``$\circ$'' denotes the composition of maps, and we routinely treat a cotangent vector $v\in T^\ast_x{\cal M}$ as a map $v\in T_x{\cal M}\to {\mathbb R}$.

For our particular problem, the adjoint state can be reduced to an element of the Lie algebra (i.e.~for $\SU{N}$ embedded in $\mathbb{C}^n$ as a skew-symmetric traceless matrix), that follows the ODE
\begin{equation}\label{eqn:adjoint_Lie}
    \dot{A} = [Z, A] + \nabla (\nabla \cdot \dot{U}) U^{-1} - \sum_a A_a (\nabla Z^a) U^{-1}\,.
\end{equation}
Again, it is initialized as $A_a(T) = \partial_a L(U(T))$, where we have specified to the loss $L$ from \eqref{eq:L}, with $q=q_T$ the distribution at time $T$.
After solving the ODE for $A$, the gradient of the loss is computed as
\begin{equation}\label{eqn:gradient_Lie}
  \partial_\theta L(U(T)) = - \int_T^0 \qty(\langle A(t), \partial_\theta Z\rangle - \sum_{a} \partial_\theta \partial_a Z^a) \dd{t} \,.
\end{equation}
A derivation of this from the general statement above can be found in the appendix.

\subsection{Numerical Integration}
Central to implementing the continuous flows for gauge theories is to define an integration scheme that works simultaneously for Lie group and scalar-valued degrees of freedom.
The former is required for the field configurations themselves, while the latter arises for the probability densities as well as the gradients with respect to parameters in the adjoint sensitivity method.
In principle, the adjoint state additionally introduces the tangent space, which we could identify as a Lie algebra element and treat separately. However, for our implementation we treat it simply as an element of the ambient complex space.

In principle, the Lie group elements themselves could also be treated as elements of the ambient space.
However, a discrete integration scheme would inevitably result in paths leaving the target manifold.
Including projection steps could address this, but the need to reliably estimate the change of density makes this naive approach unwieldy.
We, therefore, implement a Crouch-Grossmann Runge-Kutta method \cite{crouch1993NumericalIntegration, wandelt2021GeometricIntegration}, which can be defined for any matrix group and result in paths that remain in the desired submanifold up to numerical error. The algorithm is briefly outlined with further technical details in section \ref{sec:cg} of the appendix.

\subsection{Lattice Gauge Theory Flow}

In this section, we describe in some detail our choice for the neural vector field.
In pure gauge theories on the lattice with gauge group $G$ and square lattice $V_L\cong ({\mathbb
Z}/L{\mathbb
Z})^d$ with oriented edges connecting the adjacent vertices, the field configuration consists of $U_e\in G$ for each edge $e$.
Reversing the direction of the edge gives $U_{-e}=U_e^{-1}$.

The target probability density is given by the Wilson action
\begin{equation}\label{eq:wilson-action}
    S(U) = - \frac{\beta}{N} \sum_{x\in V_L} \sum_{\mu < \nu}  \Re[\tr P_{\mu\nu}(x)]\,,
\end{equation}
where $P_{\mu\nu}(x)$ are the plaquettes, namely the product of $U$ along edges forming the smallest closed loops.
Here, $x$ denotes the initial vertex, and $\mu$ and $\nu$ are the two directions that the loop spans.
We only include one orientation by requiring $\mu < \nu$ with both $\mu$ and $\nu$ pointing in the positive direction.
An illustration is given in \figref{fig:loops} (0).

For any given $\Omega: V\to G$  we denote the corresponding gauge transformation as $\Omega \cdot U$. For instance, we have
\begin{equation}
    U_e \mapsto (\Omega \cdot U)_e = \Omega_{x_i} U_e \, \Omega_{x_f}^{-1}
\end{equation}
where $x_{i/f}$ represent the starting/end-point of the edge $e$. Note that the traces of Wilson loops, and hence the Wilson action, are invariant under the above gauge transformation.

The guiding idea for constructing our continuous flow architecture is that the gradients of gauge-invariant quantities, such as traces of loops $\W$, are gauge equivariant:
\begin{equation}
    \nabla_e \W
    \xmapsto{U \mapsto \Omega \cdot U}
    \Omega_{x_i} (\nabla_e \W) \Omega_{x_f}^{-1} \,.
\end{equation}
which is the precisely property we require of the vector field $Z_\theta(U, t) U(t)$ defining the ODE \eqref{eqn:ODE1} for our continuous normalizing flows.

In terms of the directional derivatives, i.e. transporting the above back by right multiplication to the Lie algebra, this is equivalent to
\begin{equation}
    \partial_a^e \W
    \mapsto
    \Omega_{x_i} (\partial_a^e \W) \Omega_{x_i}^{-1} \,.
\end{equation}
If the ODE is gauge equivariant, then transformation  commutes with the normalizing flow map.
Since the prior we choose (the Haar measure) is by definition gauge-invariant, the generated probability distribution will automatically be gauge-invariant, namely  $q_T(U) = q_T(\Omega \cdot  U)$, just like the target Wilson action.

We will now use the Wilson loops as building blocks of our vector field.
To build a sufficiently expressive ODE,
we must use a larger set of loops than just the plaquettes.
We denote the traces of loops of some  (to be specified below) class $(k)$ as $\W_{(k)}$.
\figref{fig:loops} shows the specific classes of loops we consider for our experiments in two dimensions.
Instead of the lattice vertices, we will use elements of the dual lattice as anchor points of the Wilson loops, denoted as $\bar{x}$ and shown in the figure as dots, as this better reflects the symmetries of the loops.
In $\W_{(k)}$ we include all spatial alignments and shifts that keep the original anchor point enclosed in the loop and leave the orientation of the loop invariant.
For instance,
for each anchor point $\bar{x}$ in two dimensions, $\W_{(0)}$  gives rise to one trace value, namely the plaquette $\W^1_{(0)} = \tr P_{01}$, while $\W_{(1)}$ and $\W_{(3)}$ each gives rise to four (corresponding to up, down, left, right).

Concretely, we build the neural ODE by taking a superposition of the vectors $\partial_a^e  \W$ with coefficients depending on gauge invariant quantities, which in particular include any element of $ \W_{(k)}$.
To allow for efficient computation of the divergence, which is needed to compute the change of density, we use a shallow architecture that first applies a neural network locally in $\bar{x}$ to a subset of $ \W$, followed by a convolution in $\bar{x}$ that introduces non-local correlations.
This way the divergence computation can be done in $\mathcal{O}(|E|)$ time, while for general gauge invariant coefficients, it would take
$\mathcal{O}(|E|^2)$ time.
See section \ref{sec:div-cost} of the appendix.
Writing, as before, $\dot{U_e} = \sum_a Z^a_e(t,U) T_a U_e$,
we define
\begin{equation} \label{eq:model-general}
 Z^a_e(t,U)  =
    \sum_{k, \bar{x}}
    \qty(\partial_a^e {\W}^k(\bar{x}) )
    \Lambda^{k}_{\bar{x}} (t, {\W}) \,,
\end{equation}
with $\Lambda(t,  \W)$ a time-dependent convolution of features obtained from $ \W$ by locally applying a neural network at each anchor point $\bar{y}$:
\begin{equation}
    \Lambda^{k}_{\bar{x}} = \hat{\Lambda}^{k}(t) +
    \sum_{j,\bar{y}} C^{kj}_{\bar{x}\bar{y}}(t) \, g^{j}_{\bar{y}}\qty(t, \W(\bar{y})) \,,
\end{equation}
where $C(t)$ is a time-dependent convolution kernel, $\hat{\Lambda}^{k}(t)$ are field-independent learnable parameters, and $g$ represents a neural network.

We found that this parametrization is expressive while still allowing for a tractable divergence computation, as required for the change of density computation.
Further details of the architecture used in our experiments can be found in section \ref{sec:lattice-experiments}, where it is applied to two-dimensional pure gauge theories.

\section{Numerical Experiments}

\subsection{Single-Variable Conjugation Equivariant Flows}

\begin{figure}[htb]
    \centering
    \includegraphics[width=\linewidth]{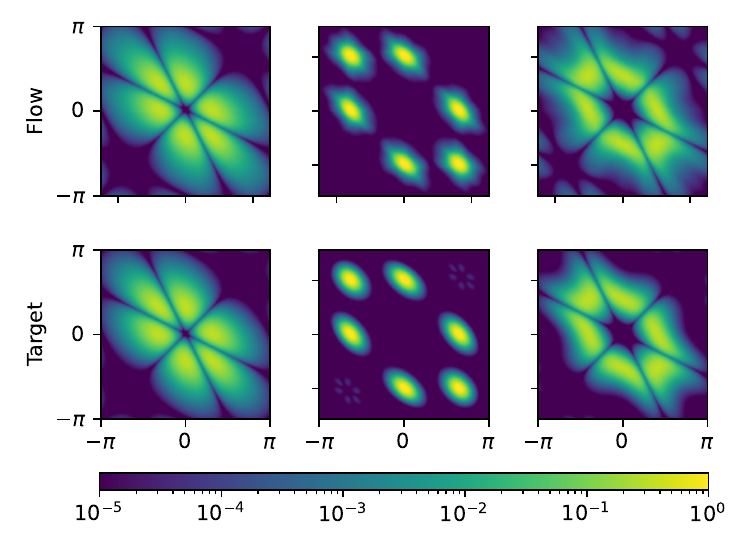}
    \vspace{-20pt}
    \caption{Comparison of the $\SU{3}$ densities at $\beta=9$ in terms of angular coordinates for the three conjugation equivariant targets listed in table \ref{tab:coefficients} and the respectively trained continuous normalizing flows.}
    \label{fig:single-densities}
\end{figure}

To highlight the flexibility of continuous flows and their expressive power, we first repeat the experiment in \cite{boyda2021SamplingUsing} and use flows to learn a conjugation-invariant density on a single copy of $\SU{N}$.
Specifically, we use three different families of target actions of the form
\begin{equation} \label{eq:conj-target}
    S^{(i)}(U) =
    - \frac{\beta}{N} \Re \tr \qty(
        c_n^{(i)} U^n
    )
\end{equation}
with the sets of values $c^{(i)}$ as shown in table \ref{tab:coefficients}.

\begin{table}[h]
  \caption{\label{tab:coefficients}%
  Table of coefficients $c{(i)}$ for the conjugation equivariant target actions of equation \eqref{eq:conj-target}.
  These are the same values as used in \cite{boyda2021SamplingUsing}.
  }
\begin{ruledtabular}
\begin{tabular}{cccc}
\textrm{Target $(i)$} & $c_1^{(i)}$ & $c_2^{(i)}$ & $c_3^{(i)}$ \\[0.2em]
\colrule\\[-0.8em]
0 & 1 & 0 & 0 \\
1 & 0.17 & -0.65 & 1.22 \\
2 & 0.98 & -0.63 & -0.21 \\
\end{tabular}
\end{ruledtabular}
\end{table}

We parametrize the ODE using a time-dependent potential denoted $P_\theta$, which we learn as a neural network:
\begin{equation}
    \dot{U} = \nabla P_\theta(t, U) \,.
\end{equation}
The gradient and the Laplacian, needed for the change of density computation following \eqref{eqn:loglikelihoodODE}, are both computed using automatic differentiation, which is feasible as we only consider a single $\SU{N}$ matrix here.

To ensure that the defined flow is conjugation equivariant, we define the potential exclusively in terms of conjugation invariant quantities.
First, we construct $64$ time features by applying the $\sin$-function with individually learnable frequencies to $t$,  followed by a dense layer with $64$ features.
For $U$-dependent features, we use absolute, real, and imaginary parts of $\tr (U^n)$ for powers $n = 1 \, \ldots \, n_{\max}$.
Specifically, we use
\begin{equation}
\begin{gathered}
    |\tr (U^n)|,\, \Re \tr (U^n),\, (\Re \tr (U^n))^2,\, \\
    \Im \tr (U^n),\, (\Im \tr (U^n))^2\,.
\end{gathered}
\end{equation}
The time and $U$-dependent features are then concatenated and processed by a ResNet of width $512$ and activation function \texttt{gelu} before being projected to a single real value representing the potential $P_\theta$.

For target $c^{(1)}$ we used ResNet depth $3$ and $n_{\max}=2$, while for the other two targets we used depth $3$ and $n_{\max}=4$.

Leveraging the continuous nature of our normalizing flows, we train a single potential for each type of target and match the flow time $t$ with the target inverse temperature $\beta$ as $\beta(t) = 9 \cdot t$.
For training, we use $4$ randomly drawn end times $T$ and corresponding betas uniformly in $T \in [0.1, 1]$ and evaluate the KL loss on $1024$ samples for each.

\begin{figure}[htb]
    \centering
    \includegraphics[width=\linewidth]{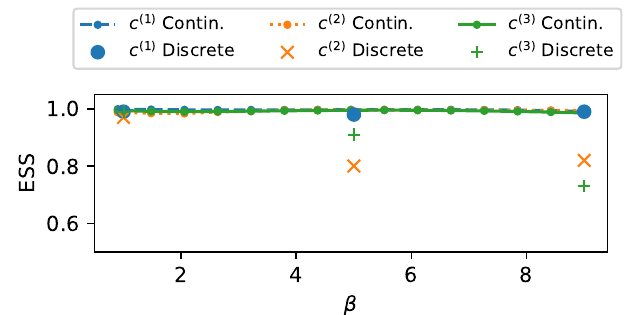}
    \caption{Effective sample size along the flow time of trained normalizing flows for each of the three families of target distributions listed in table \ref{tab:coefficients} in comparison to the ESS achieved by the individual discrete flows from \cite{boyda2021SamplingUsing}.}
    \label{fig:single-ess}
\end{figure}

The densities evaluated in angular coordinates (see appendix \ref{sec:haar} for details) of the flow shown in \figref{fig:single-densities} match the target densities visually to high accuracy.
Similarly, \figref{fig:single-ess} shows that the flows are able to approximate the target density for the whole range of target $\beta$ values to high accuracy as measured by effective sample size (ESS).

\subsection{Pure Gauge Theory in Two Dimensions}
\label{sec:lattice-experiments}

To demonstrate the feasibility of the model, and in particular the model architecture described in equation \eqref{eq:model-general}, we performed experiments on two-dimensional pure gauge theory  \eqref{eq:wilson-action} for $\SU{2}$ and $\SU{3}$, achieving state-of-the-art or better effective sample sizes.
Table \ref{tab:lat16} shows the result of training for a $16 \times 16$ lattice, comparing to previous discrete \cite{boyda2021SamplingUsing} and continuous \cite{bacchio2023LearningTrivializing} architectures.
Table \ref{tab:lat8} shows results for $\SU{3}$ on a smaller $8 \times 8$ lattice but for larger values of $\beta$.

In contrast to previous work \cite{boyda2021SamplingUsing, bacchio2023LearningTrivializing}, we have found that initially using a smaller batch size of $32$ and increasing this after some steps helps to speed up training.
Concretely, we train using a second-order Crouch-Grossmann integrator with $40$ steps in 32-bit mode until training approximately converged, then switching to 64-bit mode and a batch size of $128$.
ESS evaluation was done in each case in 64-bit mode and using $40$ steps of a third-order Crouch-Grossmann method.

We will now outline in more detail the architectures used for $\SU{2}$ and $\SU{3}$ flows, respectively.

\begin{table}[htb]
\caption{\label{tab:lat16}%
    Effective sample sizes for normalizing flows for $SU(2)$ and $SU(3)$ on $16\times16$ lattice compared to previous work. Best values in bold.
}
\begin{ruledtabular}
\begin{tabular}{lcccccc}
  & \multicolumn{2}{c}{SU(2)} & & \multicolumn{3}{c}{SU(3)} \\
  ESS [\%] &
  $\beta = 2.2$ & $\beta = 2.7$ & &
  $\beta = 5$ & $\beta = 6$ & $\beta = 8$ \\[0.2em]
  \colrule\\[-0.8em]
Continuous flow & \bf 87 & \bf 68 & & 86 & \bf 76 & \bf 23 \\
Bacchio et al \cite{bacchio2023LearningTrivializing} & -- & -- & & \bf 88 & 70 & -- \\
Boyda et al \cite{boyda2021SamplingUsing} & 80 & 56 & & 75 & 48 & --  \\
\end{tabular}
\end{ruledtabular}
\end{table}

\begin{table}[htb]
  \caption{\label{tab:lat8}%
  Effective sample sizes for normalizing flows for $SU(3)$ on $8\times8$ lattice compared to previous work. Best values in bold.
  }
\begin{ruledtabular}
\begin{tabular}{lcc}
ESS [\%] &
$\beta = 8$ &
$\beta = 12$ \\[0.2em]
\colrule\\[-0.8em]
Continuous flow & \bf 64 & \bf 27 \\
Multiscale + flow \cite{abbott2024MultiscaleNormalizing} & 35 & 13 \\
Haar + flow \cite{abbott2024MultiscaleNormalizing} & 25 & 3 \\
\end{tabular}
\end{ruledtabular}
\end{table}

\subsubsection{Flows on $\SU{2}$ lattices}

When applying the general form of the architecture in  \eqref{eq:model-general} above to the specific experiments described here, we only used the first two classes of loops, shown in red in \figref{fig:loops}, as input to the neural network that defines the superposition coefficients $\Lambda$.
Moreover, as we sum over the anchor points $\bar x$, to avoid redundancy we do not include all shifted copies of the Wilson loops in the gradients $\nabla_e \W$.
We therefore denote by ${\Tilde\W}_{(i)}$ the subset of traces ${\W}_{(i)}$ where we pick one representative of each set of shifted loops.
For example, ${\Tilde\W}_{(1)}(\bar{x})$ contains two instead of four values for each anchor point.

Let $ \W_{(i_1:i_2)}$ denote the union of $\W_{(i_1)}, \ldots, \W_{(i_2)}$.
The architecture of the neural ODE used in our experiments is then given by
\begin{equation}
   Z^a_e(t,U) =
    \sum_{k, \bar{x}}
    \qty(\partial_a^e {\Tilde\W}^k_{(0:7)}(\bar{x}) )
    \Lambda^{k}_{\bar{x}} \qty(t, {\W}_{(0:1)}) \,,
\end{equation}
where the superposition coefficients are
\begin{equation} \label{eq:superpos}
    \Lambda^{k}_{\bar{x}} = \hat{\Lambda}^{k}(t) +
    \sum_{j,\bar{y}} C^{kj}_{\bar{x}\bar{y}}(t) \, g^{j}_{\bar{y}}\qty(t, \W_{(0:1)}(\bar{y})) \,.
\end{equation}
Note that for $\SU{2}$ all traces $\W$ are real numbers, and can be directly input into conventional dense neural networks.

We parametrize $g$ as a ResNet of depth 3, width 128, and with \texttt{gelu} activation function.
The time dependence is introduced by first applying a set of 16 uniformly shifted Gaussian density functions to $t$ followed by a single dense layer with output size 16 to generate a set of features $K_{g}^l(t)$ with $l=1,\ldots,16$.

The convolution kernel is chosen to be of size $5 \times 5$. The time dependence is introduced by having $16$ copies of parameters and contracting them with $K_C^l(t)$ which is constructed as above but with independently trainable parameters in the dense layer.
The number of output features of $g$ and thus input features of $C$, indexed by $j$ in equation \eqref{eq:superpos}, are chosen to be 24.

\subsubsection{Flows on $\SU{3}$ lattices}

The architecture for $\SU{3}$ is qualitatively the same as for $\SU{2}$.
However, in this case, the traces of loops are generally complex.
For the input of the network $g$ we use $\Re \W_{(0:1)}$ and $\qty(\Im \W_{(0:1)})^2$. The latter is chosen to reflect the parity symmetry of the theory.
Similarly, We have two output channels in the convolution corresponding to the imaginary and real parts of the gradients $\partial\W$.
The final form of the architecture is then
\begin{equation}
\begin{gathered}
    Z^a_e(t,U)  =
    \sum_{k, \bar{x}}
    \Re \qty(\partial_a^e {\Tilde\W}^k_{(0:7)}(\bar{x}) )
    \Lambda^{(R) k}_{\bar{x}} \qty(t, {\W}_{(0:1)}) +\hphantom{0}
    \\
    \Im \qty(\partial_a^e {\Tilde\W}^k_{(0:7)}(\bar{x}) )
    \qty( D^{kj} \Im \W^j_{(0:1)} ) \Lambda^{(I) k}_{\bar{x}} \qty(t, {\W}_{(0:1)}) \,,
\end{gathered}
\end{equation}
where we ensure parity symmetry by introducing a dense matrix $D$, and with the two superposition factors similar to the above:
\begin{equation}
\begin{aligned}
    \Lambda^{(\ast)k}_{\bar{x}} = \hat{\Lambda}^{(\ast)k}(t) +
    \sum_{j,\bar{y}} C^{(\ast)kj}_{\bar{x}\bar{y}}(t) \, g^{j}_{\bar{y}}\qty(t, \W_{(0:1)}(\bar{y})) \,.
\end{aligned}
\end{equation}
The time dependence and network architecture are the same as those specified above for $\SU{2}$.

\section{Outlook}

In this work, we introduced a gauge-equivariant continuous normalizing flow framework for lattice gauge theories. Our method demonstrates competitive effective sample sizes (ESS) when applied to two-dimensional $\SU{2}$ and $\SU{3}$ pure gauge theories, outperforming previous discrete and continuous approaches.

In machine learning literature, the idea of guiding the continuous flow to achieve more stability and accuracy has been discussed in various works \cite{wu2020StochasticNormalizing, midgley2023FlowAnnealed}.
In our experiment on a single Lie group flow, we have briefly explored the method matching the flow time to a predetermined evolution of target density and obtained encouraging preliminary results. It would be extended to lattice gauge theories, connecting to related work of stochastic normalizing flows \cite{caselle2022StochasticNormalizingb} that have shown promising results also in higher spatial dimensions \cite{bulgarelli2024samplingsu3puregauge, Bulgarelli:2024brv}.

Scaling up normalizing flows to larger lattice sizes remains a challenge \cite{abbott2023AspectsScaling, deldebbio2021EfficientModelling}.
We believe that the broad class of continuous equivariant normalizing flows developed in this paper constitutes a promising tool for achieving large-scale flow-based lattice simulations.
To facilitate scaling, the incorporation of methods such as multiscale sampling \cite{abbott2024MultiscaleNormalizing}, transfer learning \cite{gerdes2023LearningLattice} and other hierarchical generation and transfer learning ideas should be further explored.

In this work, we have limited ourselves to simple dense networks when designing neural networks employed to construct the neural ODE. It would be interesting to explore other network architectures, including the recently proposed gauge equivariant convolutional networks \cite{favoni2022LatticeGagueEquivariant}, in the future.
Finally, it will be useful to extend the present work to include field theories with fermions.

\begin{acknowledgments}
We thank Niki Stratikopoulou for helpful discussions.
The research of MG and MC is supported by the Vidi grant (number 016.Vidi.189.182) from the Dutch Research
Council (NWO). MG was partially supported by a project that has received funding from the European Research Council (ERC) under the European Union's Horizon 2020 research and innovation programme (Grant agreement No. 864035).

\end{acknowledgments}

\appendix
\section{Adjoint Sensitivity Method}
\label{app:adjoint}
For a smooth manifold $\mathcal{M}$, we consider the ODE $\dot{z} = f_\theta(t, z) \in T_z{\mathcal M}$ where $\theta$ represents some set of trainable parameters.
Let $\Phi$ be the associated flow map
with the properties
\begin{subequations}
\begin{eqnarray}
  \dv{s} \Phi_{t, s}(z) \Big|_{s=0}&=& f_\theta(t, z) \label{eq:derivative}
  \\
  \quad \Phi_{t + s, s'} \circ \Phi_{t, s} &=& \Phi_{t, s + s'} \,, \label{eq:composition}
\end{eqnarray}
\end{subequations}
and with the boundary condition $\Phi_{t,0}(z) = z$. From the above, it is clear that the map $\Phi$ depends on the trainable parameters $\theta$ through the form of the ODE.

Given a loss function $L: \mathcal{M} \rightarrow \mathbb{R}$, we are interested in computing $\partial_\theta L(z(T))$ where $z(T) = \Phi_{0, T}(z(0))$ given a fixed initial $z(0)$.
At the final time $T$, the derivative $dL|_{z(T)}$ measures the sensitivity of the loss to small changes in the final value $z(T)$.
Formally, it is a cotangent vector $dL|_{z(T)} \in T^*_{z(T)}\mathcal{M}$ which linearly maps vectors in $T_{z(T)}\mathcal{M}$ to a real number.
To compute it, one needs to take into account the effect of changing the ODE at all earlier times $t < T$.
To this end, we define an auxiliary \textit{adjoint state} ${{a(t) \in T^*_{z(t)}\mathcal{M}}}$
defined as follows:
\begin{equation}
  a(t)
 : = d (L \circ \Phi_{t, T - t}) \big|_{z(t)}
  = dL\big|_{z(T)} \circ d\Phi_{t, T - t}\big|_{z(t)} \,.
\end{equation}
Like $dL|_{z(T)}$, the cotangent vector $a(t)$ is a linear map from a tangent vector in $T_{z(t)}\mathcal{M}$ to a real number.
It captures the sensitivity of the loss to (infinitesimally) perturbing $z(t)$ in the direction of the tangent vector, then flowing with the ODE to time $T$, and evaluating the loss.
More explicitly, this means
\begin{equation}
    a(t)(X) = d L\big|_{z(T)} (d\Phi_{t, T - t}(X))
\end{equation}
for all $X\in T_{z(t)}{\cal M}$, where
\begin{equation}
    d\Phi_{t, T - t}\big|_{z(t)}:T_{z(t)}{\cal M} \to T_{z(T)}{\cal M}
\end{equation}
is the differential map associated to
$\Phi_{t, T - t}$.
Formally, $a(t)$ is the pullback of $dL$ under the flow map $\Phi$:
$
  {{a(t) = \Phi_{t, T - t}^* \, dL}}  \,.
$

To obtain an equation for $a(t)$, we take the derivative with respect to $t$.
First, observe that
\begin{equation}
\begin{aligned}
  a(t - s)
  = d(L \circ \Phi_{t, {T-t}} \circ \Phi_{t - s, s})
  = a(t) \circ d \Phi_{t - s, s}
\end{aligned}
\end{equation}
which follows from the composition property \eqref{eq:composition} and the chain rule.
Then, using that the derivatives with respect to $z$ and $t$ commute and property \eqref{eq:derivative} we can take the time derivative to get:
\begin{equation} \label{eqn:adjoint_general_app}
\begin{aligned}
  \dot{a}(t)
  &= - \dv{s} a(t - s) \Big|_{s=0} \\
  &= - a(t) \circ d\qty( \dv{s}\Phi_{t - s, s}\Big|_{z(t-s)} )\Big|_{s=0}  \\
  &= - a(t) \circ d f_\theta\Big|_{z(t)}
\end{aligned}
\end{equation}
Note that we also use here that the tangent space is a vector space, and as such its tangent space can be identified with itself.
If we let $z(t)$ vary in ${\cal M}$ and consider $a$ as a section of the cotangent bundle, $\dot a$ then lies in its vertical bundle.

Using this, we can solve for the adjoint state by integrating the ODE \eqref{eqn:adjoint_general_app} backwards in time, starting from $z(T)$ and the initial value $a(T) = dL\big|_{z(T)}$.
Given the adjoint state, the derivative with respect to parameters is computed simultaneously as
\begin{equation}
    d_\theta L(z(T)) = - \int_T^0 a(t) \circ  d_\theta f_\theta(t, z(t)) \dd{t} \,,
\end{equation}
as explained in \cite{chen2018NeuralOrdinary}.

\subsection{Lie-Algebra Valued Adjoint State}

We will show how the explicit equation of \eqref{eqn:adjoint_Lie} for the adjoint state, represented as a Lie algebra element, follows from the general expression in equation \eqref{eqn:adjoint_general}.

We consider an ODE on $\mathcal{M} = G \times \mathbb{R}$ whose elements we denote $z=(z^{(1)}, z^{(2)})=(U, \mathcal{L})$ where $\mathcal{L} \in \mathbb{R}$ represents the log-likelihood.
In particular we may have $G = \SU{N}^{|E|}$, but to improve legibility we will suppress the lattice index on $U \in G$ below.

The ODE $\dot{z}=f_\theta$ is given by $f_\theta = (\dot{U}, \dot{\mathcal{L}})$, and using the ODE for $\mathcal{L}$ of equation \eqref{eqn:loglikelihoodODE} we have:
\begin{subequations}
\begin{gather}
    \dot{U} = Z(t, U) U = \sum_a Z^a(t,U) T_a U \,,\\
    \dot{\mathcal{L}} = - \nabla_U \cdot \dot{U}
    { = -\sum_{a} \partial_a Z^a(t, U) \,.}
\end{gather}
\end{subequations}
We integrate this forward in time from $t=0$ to $t=T$ with the initial $U(0)$ drawn from the Haar prior $q_0(U) = \rho(U)$ and $\mathcal{L}(0) = \log \rho(U(0))$.
Our loss function has the specific form
\begin{equation}
    L(U, \mathcal{L}) = S(U) \, + \, \mathcal{L} \,,
\end{equation}
which we evaluate at $z(T)=(U(T), \mathcal{L}(T))$.
Note that as our manifold ${\cal M}$ is a product space, its cotangent space at any given point $z$ can be split into an element in $T_{z^{(1)}}^\ast G$ and $T_{z^{(2)}}^\ast {\mathbb R}\cong {\mathbb R}$, and we can use the notation $d_U$  and $d_{\cal L}$ to indicate the corresponding operators in the two spaces.
Thus, we have
\begin{gather}\begin{split}
    a(T) &=  a^{(1)}(T) + a^{(2)}(T)\\
     a^{(1)}(T) &= d_U S , ~~a^{(2)}(T) =d_{\mathcal{L}} \mathcal{L} = 1.
\end{split}
\end{gather}

We have $d f_\theta = (d \dot{U}, d \dot{\mathcal{L}})$ and both vector fields only have explicit dependence on $U$, so we have $d f_\theta=d_Uf_\theta$.
To obtain the specific form of $\dot{a}(t)$ from the genral form of equation \eqref{eqn:adjoint_general_app} we compute
\begin{subequations}
\begin{gather}
    d_U \dot{U}_e = d_U (Z U) = (d_U Z) U + Z d_U U \,,
    \\
    d_U \dot{\mathcal{L}} = - d_U (\nabla_U \cdot \dot{U}) \,.
\end{gather}
\end{subequations}
This implies that the adjoint state satisfies
\begin{gather}
\begin{split}
    \dot{a} &= - a \circ d_z f_\theta  \\
   & = - a \circ (d_U (Z U), 0) + a \circ (0, d_U (\nabla_U \cdot \dot{U})) \\
   & = \dot{a}^{(1)}+\dot{a}^{(2)}\,,
\end{split}
\end{gather}
where we split $\dot a$ into
\begin{subequations}
\begin{gather}
    \dot{a}^{(1)} = - a^{(1)} \circ d_U (Z U) + a^{(2)} \circ d_U (\nabla_U \cdot \dot{U}) \,,
    \\
    \dot{a}^{(2)} = 0 \,,
\end{gather}
\end{subequations}
as the component in the vertical bundle of $T^\ast G$ and $T^{\ast}{\mathbb R}$ respectively.
Thus, given the initial condition $a^{(2)}(T) = 1$ we have $a^{(2)}(t) = 1$  for all times, and we are left with
\begin{equation}
    \dot{a}^{(1)} = - a^{(1)} \circ d_U (Z U) + d_U (\nabla_U \cdot \dot{U}) \,.
    \label{eq:adjoint-1}
\end{equation}

Recall that $a^{(1)}(t)\in T^\ast_{U(t)}{G}$. In view of the decomposition mentioned above \eqref{eqn:ODE1} and given $U(t)$, we can equivalently define $\Tilde{a}^{(1)}\in \mathfrak g^\ast$ such that for all $v\in {\mathfrak g}$ we have $\Tilde{a}^{(1)}(v) = a^{(1)}(vU) $, namely $\Tilde{a}^{(1)} =a^{(1)} \circ r_U$ where $r_U$ denotes right-multiplication by $U=U(t)$.
We thus have
\begin{equation}\label{eqn:adjoint2}
    \dv{t} \Tilde{a}^{(1)} = \dot{a}^{(1)} \circ r_U + a^{(1)} \circ r_{ZU} \,,
\end{equation}
where we have used $\dot{U} = ZU$.
Using the scalar product which gives an isomorphism ${\mathfrak g}^\ast \to {\mathfrak g}$, we can represent $\Tilde{a}^{(1)}(t)$ as a Lie algebra element $A(t)$ such that
\begin{equation}
    \Tilde{a}^{(1)}(t)(v) ={a}^{(1)}(t)(vU)= \langle A(t) , v \rangle
~~\forall \,v \in {\mathfrak g} \,.
\end{equation}
Similarly, defining the gradient $\nabla_U h \in T_U{\cal G}$ of a scalar function $h$ as
\begin{equation}
    d_U h (vU) = \langle \nabla_U h U^{-1}, v \rangle \,,
\end{equation}
the initial condition is then given by $A(T) = \nabla_U S$.
In terms of the explicit orthonormal basis $\{T_a\}$, we have $A(T)  = \sum_a (\partial_a S) T_a U$.

Following a straightforward calculation using the above relations and expressing the bilinear form as trace  \eqref{eq:scalar-product}, we finally obtain the adjoint state ODE
\begin{equation}\label{eqn:adjoint_A_appendix}
    \dot{A} = [Z, A] + \nabla_U (\nabla_U \cdot \dot{U}) U^{-1} - \sum_a A_a \nabla_U Z^a U^{-1} \,,
\end{equation}
where $[Z, A]$ is the commutator and the $a$ index is the component with respect to the Lie algebra basis.
In terms of explicit basis $\{T_a\}_a$ for ${\mathfrak g}$ and the dual basis $\{T^a\}_a$  for ${\mathfrak g}^\ast$, we have $\Tilde{a}^{(1)}(t) = \sum_a A_aT^a$ and $A(t)=\sum_a A_aT_a$.
Then \eqref{eqn:adjoint2} acting on $v\in {\mathfrak g}$ gives
\begin{gather}\notag
\sum_{a} v^a\dot{A}_a =  \\ \notag
\sum_{a,b} (-A_a v^b \partial_b Z^a +\partial_a \partial_b Z^b)- \sum_{a}A_a \langle T_a,[v,Z] \rangle ,
\end{gather}
which becomes, upon using the cyclic property of the trace,
\begin{equation}
\dot{A}_a = \sum_b(- A_b  \partial_a Z^b + \partial_a\partial_b Z^b) +    \langle T_a,[Z,A] \rangle
\end{equation}
which is precisely \eqref{eqn:adjoint_A_appendix}.
To translate the general loss integral \eqref{eqn:gradient_general}
, repeated here:
\begin{equation}
  d_\theta L(z(T)) = - \int_T^0 a(t) \circ d_\theta f_\theta(t, z(t)) \dd{t} \,,
\end{equation}
into the specialized expression \eqref{eqn:gradient_Lie}, we simply specialize to the case $d f_\theta = (d \dot{U}, d \dot{\mathcal{L}})$. With $ \dot{U}=ZU$ we get
\begin{equation}
\begin{gathered}
    a^{(1)} \circ d_\theta \dot{U}  = \langle A(t), \partial_\theta Z\rangle \\
    a^{(2)} \circ d_\theta \dot{\mathcal{L}} = - \sum_{a} \partial_\theta \partial_a Z^a
\end{gathered}
\end{equation}
and thus the loss is computed as
\begin{gather}
\begin{split}
  \partial_\theta L(z(T)) &=  \int_T^0 \left(-\langle A(t), \partial_\theta Z\rangle + \sum_{a} \partial_\theta \partial_a Z^a \right)\dd{t} \\
 &= \sum_{a}
\int_T^0 \left(- A_a(t) \partial_\theta Z^a + \partial_\theta \partial_a Z^a \right)\dd{t}
  \,.
\end{split}
\end{gather}

\section{Gradient Implementation}

In practice, we implement the adjoint sensitivity method under the assumption that $\mathcal{M}$ is embedded in $\mathbb{C}^n$ for some integer $n$, and that we represent the values as elements of this ambient space.
This is in particular the case for matrix Lie groups such as $\SU{N}$, product manifolds thereof, and the real numbers that represent the log-density of samples.
The implementation makes heavy use of JAX's \cite{jax2018github} Jacobian-vector product (JVP, also known as forward mode) and vector-Jacobian products (VJP, also known as reverse mode).
Given a function $f: \mathbb{C}^m \mapsto \mathbb{C}^n$, these generate new functions
\begin{equation}
\begin{aligned}
  \mathrm{JVP}(f):  T \mathbb{C}^m &\rightarrow T \mathbb{C}^n,&  v &\mapsto df(v) \\
  \mathrm{VJP}(f):  T^* \mathbb{C}^n &\rightarrow T^* \mathbb{C}^m,&  c &\mapsto c \circ df \,.
\end{aligned}
\end{equation}
Numerically $T^* \mathbb{C}^m \rightarrow \cong T \mathbb{C}^m \cong \mathbb{C}^m$, and the action of a dual space element $c \in T^* \mathbb{C}^m$ on $v \in T \mathbb{C}^m$ is simply the inner product $c \cdot v \in \mathbb{C}$ on $\mathbb{C}^m$.
In particular, for a real function $h: \mathbb{C}^m \mapsto \mathbb{R}$, the cotangent $dh$ can be obtained via VJP with cotangent $1 \in \mathbb{C} \cong T^* \mathbb{C}$ as input.
With these operations, the adjoint sensitivity method can thus be implemented in full generality.

\subsection{Gradients and Divergence of Loops}

For traces of loops, the gradient and second derivatives needed to compute the neural ODE and its divergence can be efficiently computed by inserting Lie group generators into the product of group elements $U_e$ labeled by the edges $e$.

If a loop has the form
\begin{equation}
    \W = \tr (U_{e_1} \ldots U_{e_i} \ldots U_{e_n}) \,,
\end{equation}
the directional derivative $\partial_a^{e_i}$ acts on $U_{e_i}$ as $\partial_a^{e_i} U_{e_i} = T_a U_{e_i}$ and $\partial_a^{e_i} U_{e_j} = 0$ for $i \neq j$.
Then for the loop we have
\begin{equation}
    \partial_a^{e_i} \W = \tr (U_{e_1} \ldots T_a U_{e_i} \ldots U_{e_n}) \,,
\end{equation}
with a sum over all occurrences if $U_{e_i}$ appears in more than one place.
For $U_{e_i}^{-1} = U_{e_i}^\dagger$, using the fact that the generators are skew-symmetric, we have $\partial_a^{e_i} U_{e_i}^\dagger = (T_a U_{e_i})^\dag = -U_{e_i}^\dagger T_a$.
Once these directional derivatives are computed, they can be propagated through the derivative of the neural network as described above.
For the divergence, one has to trace over $e$ and $a$.
The primary challenge in the implementation is to efficiently keep track of the combinatorics of where edge elements appear and to avoid redundant computations for the second derivatives arising from the divergence calculation.

\section{Computational Cost of the Divergence}
\label{sec:div-cost}

To understand the role of architecture in relation to the computational cost of the divergence, first consider a simpler scalar vector field
\begin{equation}
    \dot{\phi}_x = f_x(\phi, t) \,.
\end{equation}
Computing the divergence
\begin{equation}
    \pdv{}{\phi_x} f_x(\phi, t)
\end{equation}
for a generic $f$ would require one automatic differentiation call per lattice site value $x$, regardless of whether forward or backward mode is used.
Let $N$ be the number of such sites (i.e.~the size of the vectors $\phi$ and $f(\phi)$).
As $f$ is a function of $N$ variables, its computational cost will scale as $\mathcal O (N)$ in the best case.
A single backward or forward pass would thus also scale linearly in $N$.
Since the full divergence computation requires $N$ backward or forward passes, this leads to a total computational cost for the divergence of $\mathcal O(N^2)$.

If the analytic form of $f$ is known and chosen appropriately, it may be used to reduce this computational cost.
Before considering our gauge field architecture, we can illustrate this in the following toy example.
Let $f_x(\phi) = \sum_{y,i} M_{xy}^i\sin(\omega_i \phi_y)$.
It is clear the divergence is in fact cheaper to evaluate than the vector field itself: $\partial f_x / \partial \phi_x = \sum_i M_{xx}^i \omega_i \cos(\omega_i \phi_x)$.
This simplification arises from the linear map $M$ and the ``locally'' applied function $\sin$.
However, a black-box application of automatic differentiation for the divergence would not be able to exploit this fact, leading to a large amount of redundant computations.

Now, recall the general architecture of equation \eqref{eq:model-general}:
\begin{equation}
 Z^a_e(t,U)  =
    \sum_{k, \bar{x}}
    \qty(\partial_a^e {\W}^k(\bar{x}) )
    \Lambda^{k}_{\bar{x}} (t, {\W}) \,.
\end{equation}
Applying the chain rule, after differentiating with respect to the same link and Lie algebra component (i.e.~applying $\partial_a^e$), the divergence becomes:
\begin{equation}
\begin{gathered}
\partial_a^e Z^a_e(t,U)  = \\
    \sum_{k, \bar{x}}
    \qty(\partial_a^e\partial_a^e {\W}^k(\bar{x}) )
    \Lambda^{k}_{\bar{x}} (t, {\W})
    +\phantom{x} \\
    \sum_{k, \bar{x}}
    \sum_{j, \bar{y}}
    \qty(\partial_a^e {\W}^k(\bar{x}) )
    \qty(\partial_a^e {\W}^j(\bar{y}) )
    \pdv{}{\W^j(\bar{y})}
    \Lambda^{k}_{\bar{x}} (t, {\W})
    \,.
\end{gathered}
\end{equation}
The cost of taking the second derivative of the Wilson loops for a given edge $e$ does not increase with the lattice size, at least when  the choice of included loops is held constant.
We can thus focus on the second term.
Since not all Wilson loops contain the link $U_e$, the sums over $\bar{x}$ and $\bar{y}$ do not scale with $|E|$.
Thus, the number of derivatives $\partial \Lambda / \partial \W$ that need to be taken for any given edge remains $O(1)$.
However, if the network $\Lambda$ is a black box, these derivatives still have to be computed with automatic-differentiation for each lattice site.
Since $\Lambda$ is generically a function of all field variables, this may still scale as $\mathcal O(|E|^2)$ -- a function of $|E|$ variables evaluated $|E|$ times.

The final simplification arises from the shallow form
\begin{equation}
    \Lambda^{k}_{\bar{x}} = \hat{\Lambda}^{k}(t) +
    \sum_{l,\bar{z}} C^{kl}_{\bar{x}\bar{z}}(t) \, g^{l}_{\bar{z}}\qty(t, \W(\bar{z})) \,.
\end{equation}
Applying the derivative:
\begin{equation}
\begin{gathered}
    \pdv{}{\W^j(\bar{y})} \Lambda^{k}_{\bar{x}} = \\
    \sum_{l} C^{kl}_{\bar{x}\bar{y}}(t) \, \pdv{}{\W^j(\bar{y})} g^{l}_{\bar{y}}\qty(t, \W(\bar{y})) \,.
\end{gathered}
\end{equation}
Because the non-linearity $g$ is applied ``locally'', i.e.~its output at $\bar{z}$ only depends on the Wilson loop stack at $\bar{z}$ (which does not scale with $|E|$), the sum over $\bar{z}$ above is removed and the convolution reduces to a simple pointwise multiplication.
Thus, computing $g$ and its derivatives with respect to $\W$ does not itself scale with $|E|$.
We still have to compute this once for each link, of course, leading to a computational cost that grows linearly with $|E|$.
In contrast to a black-box vector field, however, this shows that computing the divergence for our shallow vector field architecture does not add an additional factor of $|E|$ relative to the vector field itself.

\section{Implementation of the Crouch-Grossman Runge Kutta integration method}
\label{sec:cg}

We consider ODEs on $G\times \mathbb{R}^d$, where we simultaneously evolve both real degrees of freedom $x \in \mathbb{R}^d$ and matrix Lie group degrees of freedom $Y \in G$.
To achieve this, we apply a Runge-Kutta method for the former and simultaneously a Crouch-Grossmann method \cite{crouch1993NumericalIntegration, wandelt2021GeometricIntegration} for the latter ensuring that the group structure of $Y$ is preserved
during the simultaneous time  evolution of both  variables.

In the Crouch-Grossman method,  the Lie group element evolves according to:
   \[
   Y_{n+1} = \exp(h b_s Z_s^{(n)}) \cdot \dots \cdot \exp(h b_1 Z_1^{(n)}) Y_n,
   \]
   where $h$ is the time step size, $b_i$ are the coefficients from the Butcher tableau (Runge-Kutta coefficients) and $Z_i^{(n)} = Z(x_i^{(n)}, Y_i^{(n)}) \in \mathfrak{g}$ is the Lie algebra-valued vector field at the internal stage $i$ that generally may depend on both $x_i^{(n)}$ and $Y_i^{(n)}$.

The internal stages $Y_i^{(n)} \in G$ are computed using:
\[
   Y_i^{(n)} = \exp(h a_{i,i-1} Z_{i-1}^{(n)}) \cdot \dots \cdot \exp(h a_{i,1} Z_1^{(n)}) Y_n \,,
\]
where \( a_{i,j} \) are given by the chosen Butcher tableau. While we have suppressed the explicit time-dependence in the equations above, we note that in evaluating $ Z_i^{(n)}$, we set the time to be $t_i^{(n)}=t_n+c_ih$ with $c_i=\sum_{j} a_{i,j}$.

For the real degrees of freedom $x \in \mathbb{R}^d$, we use a standard Runge-Kutta method with the same time step $h$ and the same Butcher-tableau coefficients $a_{i,j}, b_i$ as for the Lie group variables. The real degrees of freedom are updated as:
\[
   x_{n+1} = x_n + h \sum_{i=1}^s b_i f(x_i^{(n)}, Y_i^{(n)}) \,,
\]
where $f(x_i^{(n)}, Y_i^{(n)}) \in \mathbb{R}^d$ is the vector field in the ODE $\dot{x}=f(x,Y)$, governing the evolution of $x$ and $x_i$ are the internal stages computed similarly:
\[
   x_i^{(n)} = x_n + h \sum_{j=1}^{i-1} a_{i,j} f(x_j^{(n)}, Y_j^{(n)}) \,.
\]
Note that both internal stages $x_i^{(n)}$ and $Y_i^{(n)}$ are updated in parallel using their respective methods.

\section{Haar Measure and Sampling}
\label{sec:haar}

\subsection{Haar Measure for Matrix Lie Groups}

The Haar measure defines a unique invariant measure on compact Lie groups $G$ such as $\SU{N}$ \cite{duistermaat2000LieGroups}. The measure, denoted $ dU $, is
invariant under left- and right-multiplications:
\[
d(VU) = d(UV) = dU \quad \forall V, U \in G.
\]This invariance ensures that the measure is uniform across the group. With respect to this measure, the Haar prior distribution is uniform and (up to volume normalization) given as $q_0(U) = 1$.

\subsection{Sampling from the Haar Distribution}

Sampling from the Haar distribution on $ SU(N) $ can be achieved through the construction of random unitary matrices. The method involves generating random matrices with independent complex entries, followed by an orthogonalization procedure that ensures the sampled matrix lies within the $ SU(N) $ group \cite{mezzadri2007HowGeneratea}.

We start with a random complex matrix $ X \in \mathbb{C}^{N \times N} $, whose elements are drawn from a standard normal distribution.
We then apply a QR decomposition:
\[
X = QR,
\]
where $ Q \in \mathbb{C}^{N \times N} $ is a unitary matrix, and $ R \in \mathbb{C}^{N \times N} $ is upper triangular. We then extract and normalize the diagonal part of $R$ by dividing each element by its absolute value: $R' = {\rm diag}(R) / |{\rm diag}(R)|$, i.e.~$R'_{ii} = R_{ii} / |R_{ii}|$, which ensures that all elements are pure phases thus making it unitary.
We then need to divide by
\begin{equation}
    \kappa = \sqrt[N]{\det(Q)\det(R')} \,,
\end{equation}
to ensure that the matrix $ X' = Q R' / \kappa $ has a determinant of 1, so finally we have $ X' \in SU(N) $. The resulting matrix $ X' $ is then uniformly distributed according to the Haar measure on $ SU(N) $.

\subsection{Haar Measure for $ SU(3) $}
For the specific case of $ SU(3) $, the Haar measure can be expressed in terms of angular coordinates that parametrize the eigenvalues of the matrix. The maximal torus of $ SU(3) $ corresponds to diagonal matrices with phases $ \theta_1 $, $ \theta_2 $, and $ \theta_3 $, satisfying $\sum_{i=1}^3\theta_i=0$ and ensuring the  unit determinant condition.

The Haar measure on $ SU(3) $ in terms of these angular variables is given by \cite{bump2004WeylIntegration}:
\begin{equation}
    dU = d\theta_1 \, d\theta_2 \, \prod_{i<j} \left| e^{i\theta_i} - e^{i\theta_j} \right|^2 \,.
\end{equation}

\section{Markov Chain Monte Carlo \& Observables}

To verify the implementation and successful training of normalizing flows, we generated asymptotically unbiased samples using an independent Metropolis-Hastings Markov chain with the trained normalizing flows as independent proposal distributions.
Denoting the learned flow distribution as $q(U)$ and the target distribution as $p(U)$, one samples proposals $U' \sim q$ which are accepted, $U^{(i)} = U'$, with probability
\begin{equation}
    \min\left(
        1,
        \frac{
            q(U^{(i-1)}) p(U')
        }{
            q(U')p(U^{(i-1)})
        } \,
    \right)
\end{equation}
and otherwise rejected, leading to a repeat of the previous sample $U^{(i)} = U^{(i-1)}$.

The acceptance rates over a sequence of this Markov chain gives an alternative quality of fit measure for the generative models to the ESS used in the main text. Table \ref{tab:ar} gives an overview of these values, which are generally in agreement with the ESS.

\begin{table}[htb]
\caption{\label{tab:ar}%
    Acceptance rates (in $\%$) for the continuous normalizing flows trained for different gauge groups and lattice sizes using an independent Metropolis-Hastings Markov chain of length $20480$ for all configurations of tables \ref{tab:lat16} and \ref{tab:lat8}.
}
\begin{ruledtabular}
\begin{tabular}{ccccccc}
    \multicolumn{2}{c}{SU(2), L=16} & \multicolumn{3}{c}{SU(3), L=16} & \multicolumn{2}{c}{SU(3), L=8} \\
    $\beta = 2.2$ & $\beta = 2.7$ &
    $\beta = 5$ & $\beta = 6$ & $\beta = 8$ &
    $\beta = 8$ & $\beta = 12$ \\[0.2em]
\colrule\\[-0.8em]
84 & 82 & 78 & 71 & 38 & 65 & 32 \\
\end{tabular}
\end{ruledtabular}
\end{table}

\begin{figure}[htb]
    \centering
    \includegraphics[width=\linewidth]{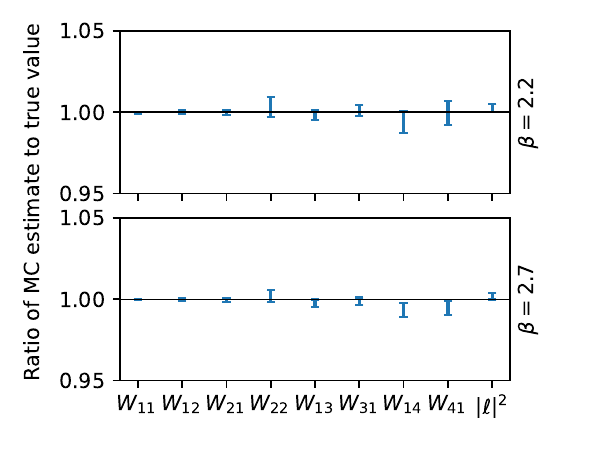}
    \vspace{-20pt}
    \caption{Ratio of Monte Carlo estimates of observables to analytic values for flows on $\SU{2}$ on a $16 \times 16$ lattice. The trained flows of table \ref{tab:lat16} are used as proposal distribution for an independent Metropolis-Hastings Markov chain of length $20480$, and errors shown as vertical bars are estimated using a bootstrap method.}
    \label{fig:observables-su2}
\end{figure}

\begin{figure}[htb]
    \centering
    \includegraphics[width=\linewidth]{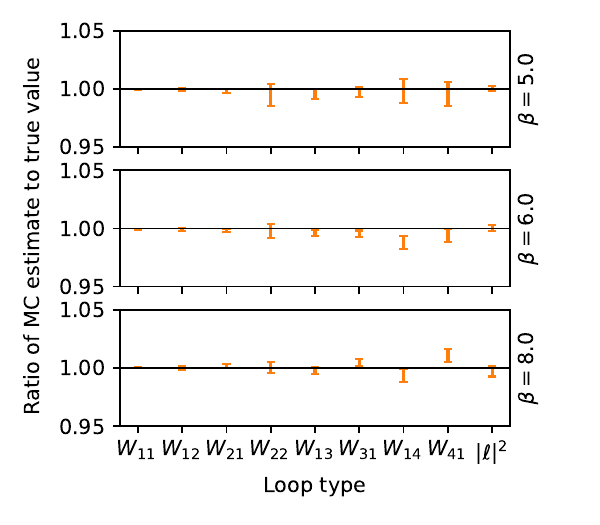}
    \vspace{-20pt}
    \caption{Ratio of Monte Carlo estimates of observables to analytic values for flows on $\SU{3}$ on a $16 \times 16$ lattice. The trained flows of table \ref{tab:lat16} are used as proposal distribution for an independent Metropolis-Hastings Markov chain of length $20480$, and errors shown as vertical bars are estimated using a bootstrap method.}
    \label{fig:observables-su3}
\end{figure}

Figures \ref{fig:observables-su2} and \ref{fig:observables-su3} show the ratios of observables estimated using the same Markov chains compared to analytically known values \cite{Migdal:1975zg,Gross:1993hu} for $\SU{2}$ and $\SU{3}$, respectively.
Here, $W_{l_1 l_2}$ denote the expectation values of the real parts of the traces of rectangular Wilson loops with edge lengths $l_1$ and $l_2$. In addition, $|\ell|^2$ is the zero-separation two-point function of Polyakov loops
\begin{equation}
    \ell(x_2) = \tr \prod_{t} U_1(x=(t,x_2)) \,,
\end{equation}
identifying axis-1 of the lattice with Euclidean time.
Then $|\ell|^2=l^*(x)l(x)$ is independent of $x$ due to translational symmetry.

\clearpage

\bibliography{main}

\end{document}